# Unseeded Large Scale PIV measurements accounting for capillary-gravity waves phase speed

**Benetazzo[1], A., Gamba[2], M., Barbariol[1], F.**

[1]{Institute of Marine Sciences (ISMAR) – National Research Council (CNR), Venice, Italy}

[2]{Autorità di Bacino dei fiumi Isonzo, Tagliamento, Livenza, Piave, Brenta-Bacchiglione, Venice, Italy}

Correspondence to: A. Benetazzo (alvise.benetazzo@ve.ismar.cnr.it)

**Abstract**

Large Scale Particle Image Velocimetry (LSPIV) is widely recognized as a reliable method to measure water surface velocity field in open channels and rivers. LSPIV technique is based on a camera view that frames the water surface in a sequence, and image-processing methods to compute water surface displacements between consecutive frames. Using LSPIV, high flow velocities, as for example flood conditions, were accurately measured, whereas determinations of low flow velocities is more challenging, especially in absence of floating seeding transported by the flow velocity. In fact, in unseeded conditions, typical surface features dynamics must be taken into account: besides surface structures convected by the current, capillary-gravity waves travel in all directions, with their own dynamics. Discrimination between all these phenomena is here discussed, providing a new method to distinguish and to correct unseeded LSPIV measurements associated with wavy structures, accounting for their phase speed magnitude and direction. This has been done measuring wavenumber vectors by using the same images exploited for PIV analysis. All measurements are performed without any flow seeding and in total absence of suspended materials, using the specular reflection of the water surface as a key marker. Results obtained at low-flow regimes in a straight concrete-made rectangular-section channel and in a river are satisfying, especially if compared to those obtained from classic LSPIV application without discrimination and correction. Moreover, a novel simple and safe procedure to orthorectify images is here presented.

**Keywords:** PIV, unseeded LSPIV, capillary-gravity waves, clustering, image orthorectification.

# 1 Introduction

Water surface geometry and motion measurements in natural environments have always been a challenge for several reasons ([3], [6]). While laboratory setups allow to operate in optimal measurement conditions, to work in safety and to protect instrumentation from possible damages, flow measurements (e.g. velocity and discharge) in riverine environments, put some main issues as the difficulty and risk of reaching the flow field when using instruments that need to be in contact with water, and then the risk of damaging them because of possible collision with suspended materials carried with the current. This is especially true in flood conditions, the most interesting in discharge monitoring, with unavoidable consequences on measurement accuracy. Moreover, most of the traditional instruments used to measure velocity are capable of 1D restitution in time, except for Doppler radars and few other instruments of recent application [10].

For these reasons, a typical laboratory measurement technique as PIV (Particle Image Velocimetry) is, has been applied since the mid-1990s to natural flow fields ([2], [13], [14]) to answer some of the most critical topics in the measurement of river flow fields and discharge. The main advantage of Large Scale PIV (LSPIV) is that, using a simple commercial video-camera, the operator can measure 2D flow field for a long time in a remote position, without being in contact with water and far from danger sources. So, it differs from most of the traditional flow instruments as it is a "completely non-intrusive" method.

Nevertheless, as LSPIV principle is based on recognizing and following particles or any other distinguishable structures or patterns on the water surface, the presence of a surface tracer, naturally or artificially added, is needed. Many authors have validated the capability of artificially seeded LSPIV ( [**1**], [9]) to measure the surface velocity, and so the discharge, in rivers and open channels, especially in high flow condition [20]. Results show that LSPIV measurements were made within acceptable accuracy, if compared with traditional instrumentations ([9], [21], [24]). LSPIV performance was tested also in naturally seeded conditions, using either foam or bubbles as flow tracer [12] or the specular reflection formed by incident light interacting with the free-surface deformations ([11], [15], [26]). Such deformations, as claimed in [30], could be of different genesis and then follow different dynamics.

Applying unseeded LSPIV in a straight concrete-made rectangular-section channel and in a riverine environment, who is writing found evidence that what LSPIV recognized on the free-surface was not all convected downstream by current velocity but, as stated in [30], two main moving structures could be found:

- surface deformations that travel with current velocity, composed principally of those free-surface deformations generated by large-scale turbulence (like vortexes);
- sequences of crests of capillary-gravity waves, of turbulent genesis too, with their own dynamics, influenced by current field.

Both of them can be present on the water surface of a river but while the former is typical of "fast" velocity flows, the latter dominates in "low" velocity flows.

The aim of the study reported in this paper was to develop an unseeded LSPIV system capable of distinguishing these two components of surface dynamics to get the only pure current contribution from the two moving structures cited. In particular, to get it from capillary-gravity waves one a correction is needed. The risk, not taking this aspect into account, is of giving an incorrect estimate of the velocity field, instantaneously and in a time and space average.

Other authors faced this problem [25], but with different approaches with respect to the one herein presented. Some hypotheses, described in detail and validated in Chapter 2 and Chapter 3, are made:

- the assumption that surface reflections sequence follows the shape of surface deformations ([5], [22], [23]) (*hyp1*);
- the horizontality of the flow (*hyp2*);
- the assumption that along current main direction, LSPIV only recognizes capillary-gravity waves traveling upstream (*hyp3*).

Being capable of distinguishing between these two contributions, by exploiting unseeded LSPIV, a wide current speed range measurement is allowed, from high flow regimes where large-scale turbulence structures convected by the main flow prevail, to the lowest ones, where unseeded LSPIV recognizes mostly capillary-gravity waves traveling with their own celerity on the free-surface. This paper focuses on the latter, because the former has already been validated by many authors ( [**1**], [2], [9], [12], [13], [14], [15], [20], [26]). Besides this, a novel procedure to orthorectify images is proposed and some case studies are described.

## 2 Hardware equipment and image pre-processing

Several experiments were conducted in a straight concrete-made rectangular-section channel and in a river with the intention to improve unseeded LSPIV performance by taking into account capillary-gravity (referred throughout this paper as c-g) waves dynamics, so extending applicability of this measurement method to lower hydraulic regimes too, where this surface phenomenon prevails.

The experimental setup was the typical one used in classic LSPIV application [26]. RGB images (1920x1080 pixels at 25 frames per second) were recorded with a CANON® Legria HFS10 High-Definition video-camera (1/2.6" CMOS sensor, 6.4-64.0 mm focal length, manual focus, PAL standard, MPEG-4 AVC/H.264 video compression) mounted on a telescopic MANFROTTO® tripod, 2 meters high. From RGB layers, the blue one was selected to create 8-bit images input for PIV algorithm.

Velocity measurements in the straight concrete-made rectangular-section channel were done from a catwalk crossing it while those in riverine environment (in Ponte San Nicolò, Italy, on Bacchiglione River) from a bridge to frame the most of the river cross-section (

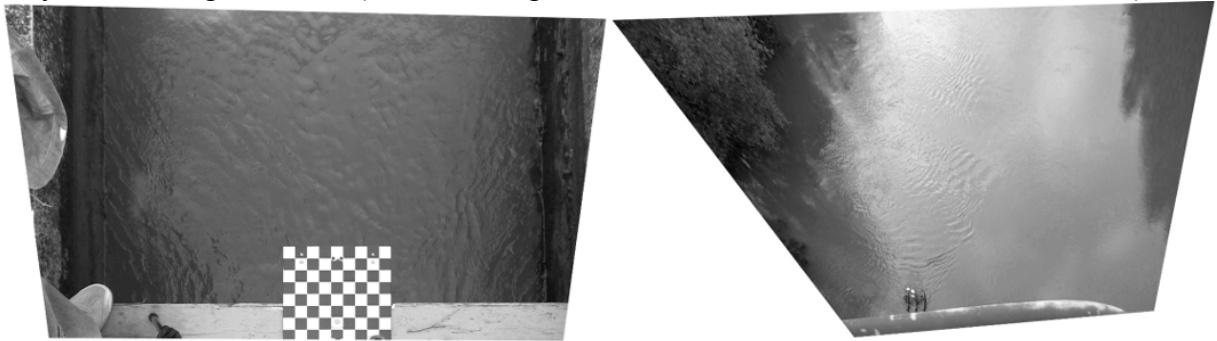

Fig. 1). Images were recorded under natural lighting, trying to minimize glare and shadow areas caused by the sun and by the presence of vegetation (in both channel and river experiments) and by the concrete side walls of the channel. No artificial seeding of the flow surface was made. Natural tracer consisted of light intensity variation associated with free-surface deformations, mainly with their slope, whose probability distribution can be measured from this intensity variation (named "hyp1") ([22], [23]). These deformations are linked with turbulence or generated by flow interaction with vegetation (also in the channel case) and with the side walls or banks.

Before being analyzed by PIV algorithm, images time sequence must be pre-processed to (i) correct angular aberration of real lens [8] and to (ii) orthorectify images to set image axes parallel and coincident with mean water surface lying position.

To correct angular aberration and to orthorectify images, a 27 cm x 27 cm black and white flat chessboard (3 cm x 3 cm squares) was used. A novel procedure for images orthorectification is herein proposed. The chessboard, kept horizontal thanks to a couple of high-precision levels and supposed parallel to the free-surface under the hypothesis of horizontal flow, is used to transfer images from camera to chessboard reference frame, through a 2D projective mapping [17]. The needed 8 degrees of freedom can be determined from the coordinates of at least four points on the chessboard in the camera reference frame ($i_k$, $j_k$) and in the chessboard one ($x_k$, $y_k$). To compute the forward mapping matrix, eight equations in the eight unknowns a, b, c, d, e, f, g, h:

$$x_k = \frac{ai_k + bj_k + c}{gi_k + hj_k + 1}, \quad y_k = \frac{di_k + ej_k + f}{gi_k + hj_k + 1} \text{ for } k = 1, 2, 3, 4. \tag{1}$$

or the equivalent 8x8 linear system (2) have to be solved.

$$\begin{bmatrix} i_1 & j_1 & 1 & 0 & 0 & 0 & -i_1 x_1 & -j_1 x_1 \\ i_2 & j_2 & 1 & 0 & 0 & 0 & -i_2 x_2 & -j_2 x_2 \\ i_3 & j_3 & 1 & 0 & 0 & 0 & -i_3 x_3 & -j_3 x_3 \\ i_4 & j_4 & 1 & 0 & 0 & 0 & -i_4 x_4 & -j_4 x_4 \\ 0 & 0 & 0 & i_1 & j_1 & 1 & -i_1 y_1 & -j_1 y_1 \\ 0 & 0 & 0 & i_2 & j_2 & 1 & -i_2 y_2 & -j_2 y_2 \\ 0 & 0 & 0 & i_3 & j_3 & 1 & -i_3 y_3 & -j_3 y_3 \\ 0 & 0 & 0 & i_4 & j_4 & 1 & -i_4 y_4 & -j_4 y_4 \end{bmatrix} \begin{bmatrix} a \\ b \\ c \\ d \\ e \\ f \\ g \\ h \end{bmatrix} = \begin{bmatrix} x_1 \\ x_2 \\ x_3 \\ x_4 \\ y_1 \\ y_2 \\ y_3 \\ y_4 \end{bmatrix} \tag{2}$$

The inverse transformation that allows setting camera axes parallel to chessboard axes is:

$$i_k = \frac{Ax_k + By_k + C}{Gx_k + Hy_k + I}, \quad j_k = \frac{Dx_k + Ey_k + F}{Gx_k + Hy_k + I} \text{ for } k = 1, 2, 3, 4. \tag{3}$$

where

$$\begin{bmatrix} A & D & G \\ B & E & H \\ C & F & I \end{bmatrix} = \begin{bmatrix} ei - fh & fg - di & dh - eg \\ ch - bi & ai - cg & bg - ah \\ bf - ce & cd - af & ae - db \end{bmatrix} \tag{4}$$

This 8x8 linear system can be solved for *A, B, C, D, E, F, G, H*.

After this step, images are transformed to fit the plane passing through the chessboard. To scale images axes in order to put them onto the water surface mean plane, a known linear dimension on the water is sufficient (i.e. the side by side cross-section width, the length of a buoyant object on the surface framed by the camera and linked with a rope, etc …) to pass from chessboard reference frame to water surface reference frame, linearly scaling both image axes with $r = l_m/l_p$, where $l_m$ is the linear dimension of the object in the free-surface reference frame and $l_p$ is the linear dimension in the chessboard reference frame (Fig. 2). To increase robustness of transformation, the value of *r* is averaged over more than an image (in our experiments, 20 images).

The adopted technique doesn't require the geo-positioning of ground control points (GCPs) and allows the operator to remain in safety on the camera site as it is only necessary that the chessboard is kept parallel to water surface. Moreover, the chessboard needs to be framed only for few seconds and the known object on the surface too. Beside this, working under the hypothesis of horizontal flow (named "*hyp2*") means that this procedure should be used in rivers where it is reasonable to assume that the free-surface slope along the mean current direction is approximately zero. However, the error committed under this assumption can be estimated as follows: if the water surface slope $i = tan(\gamma)$ is neglected, where $\gamma$ is the inclination angle of water surface with respect to the horizontal plane, LSPIV measures a *ds* displacement in time interval *dt* (and so $v_0 = ds/dt$) instead of a *ds/cos*($\gamma$) displacement (and so $v_i = ds/(cos(\gamma)dt) = (i\ ds)/(sen(\gamma)dt)$) in the same *dt*. In a river with a water surface slope $i = 10^{-2}$ (i.e. $\gamma \approx 0.6°$, a maximum value for almost all river discharge conditions), the error committed on velocity measurements is in the order of $10^{-5}$ m/s.

## 3   PIV principles and velocity values correction

The experimental setup described has been employed to validate the application of unseeded LSPIV in those flow regimes where c-g waves dynamics influences the instantaneous water surface elevation, velocity and pattern. Under *hyp1* and *hyp2*, flow velocity and image pattern wavenumber components along orthogonal *x* and *y* image axes, $(U_{m,x}, U_{m,y})$ and $(k_x, k_y)$ respectively, are measured through image processing.

## 3.1 Principles of LSPIV algorithm and wavenumber measurement

Particle Image Velocimetry (PIV), and so Large Scale PIV (LSPIV), is based on the displacement measurement of a flow surface pattern (particles, objects, a sequence of surface waves crests, etc…) framed in the Interrogation Area (IA) in the image at time $t$, in the Search Area (SA) in the image at time $t + dt$ (Fig. 3), where $dt = 1/25$ s for images acquired.

IA and SA are taken from two subsequent images and centered on the same pixel coordinates. IA ($MX$ x $MY$ pixels) should be large enough to contain (i) the average displacement of a particle during $dt$ and (ii) a c-g wave length (at least two consecutive crests). Moreover, IA has to be small enough to avoid ambiguity in the recognition because too many possible targets are framed. On the other side, SA has to be equal or larger than IA. This is left to the experience of the operator and to a preliminary analysis on the recorded images.

Recognition of IA movement is performed through a normalized cross-correlation between IA and SA:

$$R_{ab} = \frac{\sum_{x=1}^{MX}\sum_{y=1}^{MY}[(a_{xy}-\bar{a}_{xy})(b_{xy}-\bar{b}_{xy})]}{[\sum_{x=1}^{MX}\sum_{y=1}^{MY}(a_{xy}-\bar{a}_{xy})^2 \sum_{x=1}^{MX}\sum_{y=1}^{MY}(b_{xy}-\bar{b}_{xy})^2]^{0.5}} \tag{5}$$

where $a_{xy}$ are the gray values of IA and $b_{xy}$ are the gray values of SA sub-image of same size of IA. Correlating all possible so defined SA sub-images, $R_{ab}$ is a 2-D map and the pixel position of its maximum is considered to be the displacement of IA in SA. To speed up the process, cross-correlation is made in the Fourier domain ([4]).

A threshold on maximum cross-correlation coefficient $R_{ab}$ is used to discard the lowest ones that in general correspond to bad recognition or to noisy IA or SA. Another threshold is assigned on signal-to-noise ratio, defined as the ratio between the maximum normalized cross-cor coefficient and the mean value of the other cross-correlation coefficients calculated.

After correlation, displacement accuracy is improved performing a two-dimensional Gaussian regression for sub-pixel displacement evaluation. With this procedure and this kind of interpolator, pixel locking is avoided [27] and displacement accuracy is increased.

To define 2D texture of objects that ride water surface, a calculation of wavenumbers of gray-values patterns in IA is made. This measurement is fundamental (as explained in Paragraphs 3.2 and 3.3) to identify dynamics of c-g waves moving on the surface. Following this approach, c-g waves peak wavenumber is estimated calculating the 2D Fourier Transform of IA [18], with zero-padding to increase angular and spatial resolution (Fig. 4). Only the wavenumber vector direction

$$\alpha = \arctan(k_y / k_x) \tag{6}$$

with respect to the *x*-axis, and its magnitude $|\mathbf{k}| = k = \sqrt{k_x^2 + k_y^2}$ are calculated, not the sense of propagation along the known direction. According to [19], the directional resolution is calculated through

$$\Delta\alpha = 57.32° /(k / \Delta k) \tag{7}$$

where $\Delta k$ is the wavenumber resolution, depending on pixel footprint and 2D Fourier transform dimension. Peak wavenumber is considered to be linked to image gray values pattern that mostly influences the PIV measured displacement.

## 3.2  Dynamics of free-surface deformations

It was observed that applying unseeded LSPIV in a straight concrete-made rectangular-section channel and in riverine environments, PIV algorithm follows surface patterns that move with different velocities. This fact can be seen for example in Fig. 5 where some results of unseeded LSPIV measurements are shown: streamwise velocity component $U_{m,y}$ values measured at different time and at the same analysis point on the water surface are herein plotted versus corresponding peak wavenumber component $k_y$ values. In Figure, mean velocity component measured in the same test with a reference instrument is also shown.

What can be observed is that LSPIV measures only some velocity values next to the reference (approximately 0.9 m/s). Three separate velocity groups can be identified:
- *Group a*: velocity values approximately coincident with the reference velocity. These are considered to be linked to turbulent structures (i.e. vortexes) convected by the surface flow;
- *Group b*: velocity values lower than the reference velocity. These are considered to be linked to turbulent-genesis structures (i.e. waves) convected by the surface flow but with their own dynamics that affect displacement detected by LSPIV technique;
- *Group c*: possible noise.

C-g waves propagation over the framed surface was responsible for what shown in Fig. 5 and this is confirmed by [30] where the authors clearly state that the wrinkled free-surface above a turbulent channel flow mainly consists of random c-g waves that spawn from everywhere and travel in all directions across the surface (Fig. 6).

If one wants to use velocity values linked to c-g waves traveling on the surface (*Group b* in Fig. 5) for flow velocity measurements, these velocities must be corrected to take into account wave dynamics. To do this, the Doppler shifted dispersion relation of c-g sinusoidal waves on a current must be introduced. For a fixed reference frame, it can be written as:

$$\omega_0(\vec{k}_0) = \omega(\vec{k}) \pm \vec{U} \cdot \vec{k} \tag{8}$$

where $\omega = 2\pi/T$ is the angular frequency [rad/s] of a wave with period $T$ [s], $\vec{k}$, such that $|\vec{k}| = 2\pi/\lambda$, is the wavenumber vector [rad/m], $\lambda$ is the wave length [m], $\vec{U}$ is the pure current speed vector [m/s] and subscript "$_0$" is for wave features in still water [28]. From $\omega = \vec{c} \cdot \vec{k}$, the previous relation can be rewritten as

$$\vec{c}_0 \cdot \vec{k}_0 = \vec{c} \cdot \vec{k} \pm \vec{U} \cdot \vec{k} = (\vec{c} \pm \vec{U}) \cdot \vec{k} = \vec{U}_m \cdot \vec{k} \tag{9}$$

where $\vec{c}$ is the wave phase speed [m/s] and $\vec{U}_m = (U_{m,x}, U_{m,y})$ is velocity measured by LSPIV [m/s], vectorial sum of pure current speed $\vec{U} = (u, v)$ and phase speed $\vec{c} = (c_x, c_y)$ (their magnitudes are added when waves and current travel in the same sense and subtracted when their senses are opposed). Neglecting air density, whose value is negligible with respect to water density, c-g waves phase speed can be written as

$$|\vec{c}| = \sqrt{g/|\vec{k}| + \sigma|\vec{k}|/\rho} \text{ (strictly valid in deep water, i.e. } kd \to \infty\text{, with } d : \text{water depth m])} \tag{10}$$

where $g$ is gravity acceleration [m/s$^2$], $\sigma$ is water surface tension [N/m] and $\rho$ is water density [kg/m$^3$]. This function shows a minimum (Fig. 7) for $\lambda_{min} = 2\pi/|\vec{k}| = 0.017$ m that gives $|\vec{c}_{min}| = 0.23$ m/s [29]. In all written formulae, $\vec{k}$ is referred to be the peak wavenumber, i.e. the one of the wave that shows the wider gray-values span in the IA.

Besides Doppler shift effect, according to linear wave theory, waves traveling on a current change their wave height (shoaling effect), wavelength and consequently their wavenumber, phase speed and slope, depending if they are traveling upstream (shortening and steepening effect) or downstream (stretching and flattening effect).

Moreover, it can be demonstrated that c-g waves slope under the effect of current steepening is much higher than under the effect of current flattening, with respect to the same still water wavelength $\lambda_0$, still water wave height $H_0$ [m] and current speed magnitude $|\vec{U}|$ [m/s]. An example is reported in Fig. 8, where wave slope $s$, changed because of shoaling and Doppler shift effects (steepened $s\text{-} = H\text{-}/\lambda\text{-}$, with current and wave senses and directions opposing and flattened $s+ = H+/\lambda+$, with current and wave senses and directions concordant), is normalized with respect to the still water wave slope $s_0$, and plotted versus different still water wavelength $\lambda_0$. Result wanted is highlighted by $s\text{-} / s+$ ratio. In Figure, wave slope is calculated without wave-breaking effect that limits it, having single out $\lambda$ from (9) and (10), and calculated $H$ according to [7] under shoaling effect.

Taking into account the effect of steepening and flattening cited, under *hyp1*, unseeded LSPIV algorithm can better recognize those waves whose slope is increased by the mentioned effect, i.e. the c-g waves traveling upstream. According to this, it was done the hypothesis (named "*hyp3*") that the phase speed component $c_y$, along the streamwise direction, is always opposed to the pure current component $v$. This is confirmed by our experimental evidence: PIV algorithm has never calculated velocity values larger than the reference mean current speed (as seen in example in Fig. 5).

## 3.3 Velocity clustering and correction according to capillary-gravity waves phase speed

As described in Paragraph 3.2, when performing unseeded LSPIV in environments where water surface dynamics is strongly influenced by c-g waves (i.e. low velocity flows) it becomes important to be able to distinguish between those velocity measurements that refer to pure current dynamics (*Group a* in Fig. 5) and those referring to waves dynamics (*Group b* in Fig. 5). The risk, not taking this aspect into account, is to underestimate the velocity field, instantaneously and in a time and space average (as stated in *hyp3*, i.e. for c-g waves LSPIV measures a velocity lower than the pure current velocity).

The presented method aims to discriminate between these two phenomena in LSPIV measurements and then, under the three hypotheses made, to correct those measured velocity values affected by c-g waves dynamics, according to their phase speed.

The distinction between the two groups cited is made running an algorithm of Cluster analysis [16] on the velocity values data set, measured by LSPIV at each point of analysis for all time steps. The "*k-means*" algorithm has been applied to the measured streamwise velocity component $U_{m,y}$, in order to find a specified number of clusters, following a principle of minimum distance. In the spanwise direction (orthogonal to *y*), clustering algorithm wasn't run: velocity values measured by LSPIV are assigned to the clusters referring to the corresponding streamwise component values, e.g. if $U_{m,y}$ at generic time *t* was assigned to the pure-current cluster, then $U_{m,x}$ is marked as pure-current too.

Since the clustering is made in a velocity – velocity Cartesian plan, the distance criterion used is irrelevant (the Squared Euclidean distance is implemented). In order to detect the two wanted velocity sets (structures convected by mean current velocity and c-g waves with their dynamics), the number of clusters has been set to 3 (as in Fig. 9 example). The third one is useful to identify possible noise, and eventually it could be merged with one of the other clusters.

After clustering along mean current direction (say *y*), measured velocities are classified as follows (Fig. 9):

- cluster with maximum velocity centroid is classified as "pure current" (*Cluster a*). It is true under *hyp 3*;
- maximum c-g waves phase speed (10) is calculated (*Threshold 1*) within possible wavelength range in IA;
- if velocity centroid of one of the two other clusters (say *Cluster b* and *Cluster c*) differs from *Cluster a* more than *Threshold 1*, it is classified as noise, and then discarded (*Cluster c* in Fig. 9);
- if velocity centroid of one of the two other clusters differs from *Cluster a* less then the minimum c-g wave phase speed, i.e. 0.23 m/s (*Threshold 2*), it is considered composed of velocity oscillations around the mean value, classified as pure current and merged with *Cluster a*;
- if velocity centroid of one of the two other clusters differs from *Cluster a* less than *Threshold 1* and more than *Threshold 2*, it is classified as c-g waves riding the current (*Cluster b* in Fig. 9) and its velocity values are corrected to obtain the pure current contribution.

Along direction orthogonal to mean current direction (say *x*) correction is applied to measurements belonging to wavy structures cluster, taking measured pure-current vectors as a target, i.e. sign of correction is chosen to fit pure current contribution.

Projecting $\vec{U}_m$ along *x* and *y* axis, and under the hypothesis made, from (6), (9) and (10) the correction is the following:

$$\begin{cases} U_{m,x} = u \pm c_x \\ U_{m,y} = v - c_y \end{cases} \rightarrow \begin{cases} u = U_{m,x} \mp c_x \\ v = U_{m,y} + c_y \end{cases} \rightarrow \begin{cases} u = U_{m,x} \mp |\vec{c}| \cdot \sin(\alpha) \\ v = U_{m,y} + |\vec{c}| \cdot \cos(\alpha) \end{cases} \quad (11)$$

where *u* and *v* are the wanted pure-current components in the streamwise and spanwise direction respectively and *α* is wave propagation direction with respect to *x*-axis. The clustering-correction procedure is applied to all measurement points chosen in the image independently each other.

## 4 Results and discussion

To test unseeded LSPIV in presence of c-g waves on the water surface, results of four experiments are here reported: two low-flow regime experiments in a straight concrete-made rectangular-section channel and an experiment in riverine environment, on Bacchiglione River (Italy), with contributions from pure current and from c-g waves phase speed.

Another experiment in the cited straight concrete-made rectangular-section channel was made to test unseeded LSPIV performance at very slow velocity (pure current speed comparable to minimum c-g waves phase speed, so that most of the recognizable patterns, wave crests, travel in the opposite direction with respect to mean flow current), showing only wavy contribution. Since results of this experiment were obtained in a slightly different way from that described in the Paragraph 3.3, they are reported in a separate section.

All the experiments done are performed without any flow seeding and in total absence of suspended materials, using the specular reflection of the water surface as tracer.

## 4.1 Experiments in a straight concrete-made rectangular-section channel (Experiments 1, 2): clustering-correction LSPIV

Two different discharges were set upstream of the straight concrete-made rectangular-section channel, focusing on those flow regimes where capillary gravity waves dynamics influences PIV measured surface velocity. Experiments 1 and 2 cover a discharge range where vortexes and c-g waves displacements are clearly recognizable. To test LSPIV accuracy, simultaneous measurements of the surface velocity were made with a reference instruments: a NORTEK® "Vectrino", positioned in the center of the channel, a few centimeters below the water surface to measure 3D velocity next to it. "Vectrino" technical specifications and custom parameters used during experiments and results from "Vectrino" measurements are reported below in Table 1 and Table 2 respectively. For Experiments 1 and 2, $v$ is the streamwise velocity component, positive downstream (in Figures, from top to bottom), $u$ is spanwise velocity component (in Figures, positive from left to right). Froude numbers for streamwise component of current are calculated assuming 1 m water depth. Table 3 reports the space-average of the 60 seconds time-averaged velocities calculated by LSPIV at the analysis points for the first two experiments. Contributions of pure current, the ones from correction according to c-g waves phase speed, their averaged values and results without clustering and correction are highlighted.

Comparison with "Vectrino" measurements is done taking into account the unseeded LSPIV results obtained averaging along all measurement points in the stream direction at the center of the channel, where the reference instrument was set. Table 4 reports results from classic unseeded LSPIV (named LSPIV) and clustering-correction unseeded LSPIV (named ccLSPIV) application in the center of the channel, in front of "Vectrino".

Figures from Experiment 1 and 2 results show: (i) 60 s-mean 2D velocity vector field obtained applying classic unseeded LSPIV (without clustering and correction), (Fig. 10 and Fig. 12); (ii) 60 s-mean 2D velocity vector field obtained applying clustering-correction unseeded LSPIV, in accordance with (11), (Fig. 11 and Fig. 13); (iii) 60 s-mean streamwise averaged $v$ profiles along a cross-section for classic (LSPIV) and clustering-correction (ccLSPIV) unseeded LSPIVs (Fig. 19).

## 4.2 Experiment in riverine environment (Experiment 3)

Results from an experiment in riverine environment (Experiment 3), on Bacchiglione River in Italy, are shown (Fig. 14 and Fig. 15). Simultaneous superficial measurements performed by ARPAV ("Agenzia Regionale per la Prevenzione e Protezione Ambientale del Veneto" – Servizio Idrografico Regionale, the hydrographic section of regional agency for environmental protection in Veneto, Italy) with a SONTEK® "River Surveyor M9" (Acoustic Doppler Current Profiler) mounted on a trimaran was used to compare LSPIV measured velocity (Fig. 16). Froude number of measured velocity component $v$ is $Fr_v = 0.17$ assuming a 2.50 m average water depth over the river cross-section. $v$ is the streamwise velocity component, positive downstream (in Figures, from bottom to top), $u$ is spanwise velocity component (in Figures, positive from left to right). Results show: (i) 40 s-mean 2D velocity vector field obtained applying classic unseeded LSPIV (without clustering and correction) (Fig. 14); (ii) 40 s-mean 2D velocity vector field obtained applying clustering-correction unseeded LSPIV, in accordance with (11) (Fig. 15); (iii) comparison between reference instantaneous velocity distribution (SONTEK® "River Surveyor M9" measurements) along a river cross-section and 40 s-mean simultaneous LSPIV measurements (Fig. 16). In Fig. 16 LSPIV results (classic and clustering-correction LSPIVs) are shown in terms of velocity component $v$ averaged over all points of same $x$ coordinate of Fig. 14 e Fig. 15.

## 4.3 Experiments in a straight concrete-made rectangular-section channel (Experiment 4): correction LSPIV

A further example of the remarkable role of correction according to c-g waves phase speed in unseeded LSPIV is reported. Experiment 4 was done in the straight concrete-made rectangular-section channel to test unseeded LSPIV performance in limit conditions, when the mean current speed is comparable or even lower than the minimum c-g waves phase speed $\left|\vec{c}_{min}\right| = 0.23$ m/s. Reference measurements (using "Vectrino" with Table 1 setup) yield to velocity reported in Table 5.

Experiment 4 needs some extra considerations. First of all, since mean current speed is lower than $\left|\vec{c}_{min}\right|$, LSPIV targets (waves traveling upstream) move in the opposite sense with respect to the mean current. Secondary, clustering cannot be performed as described in Paragraph 3.3 and shown in Fig. 9, because of the lack of object convected by current (i.e. vortexes) at very slow velocity: wavy component is the unique feature present on the water surface. For these reasons, in this experiment, correction according to (11) is applied to the whole set of measured velocity components $\vec{U}_m = \left(U_{m,x}, U_{m,y}\right)$, without clustering. Results,

highlighting significance of correction, are reported in Table 6 and in Figures, showing: (i) 60 s-mean 2D velocity vector field obtained applying classic unseeded LSPIV (without correction) (Fig. 17); (ii) 60 s-mean 2D velocity vector field obtained applying correction unseeded LSPIV, in accordance with (11) (Fig. 18); (iii) 60 s-mean streamwise averaged $v$ profiles along a cross-section for classic (LSPIV) and correction (cLSPIV) unseeded LSPIVs (Fig. 19).

### 4.4 Discussion

Experiments done show the advantages of accounting for c-g waves phase speed when performing unseeded LSPIV at low flow regimes: both clustering-correction LSPIV and correction LSPIV. As a matter of fact, results obtained in the center of the test channel with velocity clustering and correction LSPIV (Table 4) are in agreement with "Vectrino" ones (Table 2), within a difference of 5% (Experiment 1) and of 2% (Experiment 2) on streamwise velocity component $v$ (Fig. 21). Moreover, vectors direction is consistent with that one typical of an open-channel flow (Fig. 11 and Fig. 13). Performing LSPIV without clustering and correction, would lead to an underestimation of vectors magnitude (Table 3), and to a consequent underestimation of about 48% (Experiment 1) and 31% (Experiment 2) on averaged current speed component $v$ in front of "Vectrino" (Table 4, Fig. 19, Fig. 21), and to less accurate estimate of the vectors field (Fig. 10 and Fig. 12).

Experiment 3 has been performed in the surface area behind the "River Surveyor M9" hull. Fig. 16 shows that, applying the correction proposed, results fit velocity distribution measured by "River Surveyor M9", whereas results of classic LSPIV differ in terms of $v$ from ADCP measurements. While Fig. 14 and Fig. 20 show that vectors tend to follow the pure current-ship waves combination direction if no correction is applied, vectors direction obtained performing clustering-correction LSPIV seems in a better agreement with vectors direction in an open channel flow (Fig. 15 and Fig. 20).

Experiment 4 results show that average velocity in the center of the channel is in agreement with "Vectrino" measurements within 14% of difference on velocity component $v$, while not applying correction to LSPIV measurements according to c-g waves phase speed would have led to an underestimation of about 82% on $v$ in front of "Vectrino" (Table 5, Table 6, Fig. 19 and Fig. 21) and to an incorrect estimate of vectors field that, following wave crests propagation, would have gone even upstream (Fig. 17 and Fig. 18).

Looking at the three experiments in the straight concrete-made rectangular-section channel (Fig. 21), one can see that for bigger discharge, and so faster current, the percentage difference between results without clustering and correction and results from "Vectrino" measurements, or alternatively between the former and results from clustering-correction unseeded LSPIV (ccLSPIV), decreases, confirming the idea that accounting for c-g waves phase speed becomes more important at low-flow regimes, especially when c-g waves on the water surface mostly define images pattern and flow velocities are close to c-g waves phase speed.

Measurement goodness of presented method has been verified using accredited comparing instruments (i.e. NORTEK® "Vectrino" and SONTEK® "River Surveyor M9") and can be identified with the percentage differences in Fig. 21 for the experiment in the straight concrete-made rectangular-section channel. Concerning errors in velocity estimate, in [21] the author pointed out a detailed list of error sources in PIV technique. PIV-related error before clustering and correction, due to $ds$ integer displacement calculation, neglecting sub-pixel improvement, was estimated for all the experiments done. This error is linked to the experimental setup (camera-water surface distance, camera features and camera axes rotation with respect to water surface reference frame) and dependent on mean current speed value. Fig. 22 reports potential exceeded error values related to the percentage cumulative number of velocity component $v$ values measured by LSPIV. Since accuracy $e_s$ is 0.5 pixels, maximum exceeded error ($err = e_s / ds$) is 50%. Sup-pixel detection is claimed to improve displacement measurement accuracy [27]. Nevertheless, sub-pixel $e_s$ value is unknown, so $err$ after sup-pixel detection is not available.

## 5  Conclusions

Four experiments have been conducted in a straight concrete-made rectangular-section channel and in riverine environment to test unseeded LSPIV performance in measuring superficial flow field at low velocity regimes using the specular reflections of the water surface as tracer. In this flow regime, surface dynamics is strongly influenced by capillary-gravity waves, whose shape is mainly recognized by LSPIV system, assuming that gray value pattern in images is linked to the wave slope. To prevent incorrect estimate of the surface velocity magnitude and incorrect estimate of velocity vectors direction, through "Cluster analysis", discrimination was done between those surface structures convected by flow current and those moving in accordance with capillary-gravity waves dispersion relation.

After discrimination has been made, under the hypothesis that waves recognized by unseeded LSPIV system travel against current, PIV velocity measurements were corrected in accordance to capillary-gravity waves phase speed $\vec{c}$, inferred from measurement of wavenumber $\vec{k}$ components. Results obtained in the first two experiments done (clustering-correction LSPIV in a straight concrete-made rectangular section channel) are promising (5% maximum difference on streamwise velocity component *v*, from NORTEK® "Vectrino" comparing measurements), specially if compared to the corresponding measurements from classic unseeded LSPIV, returned to be misleading at these low flow regimes. Clustering-correction unseeded LSPIV has also been applied with success in riverine environments, on Bacchiglione River, in Italy. Performing unseeded LSPIV in limit conditions, i.e. to surface fields with current speed comparable to the minimum capillary-gravity waves phase speed, the lack of clearly recognizable structures convected by the flow current and the predominance of wavy structures yielded to apply correction to the whole set of measured velocity components, without clustering: correction LSPIV led to satisfying results (14% difference on streamwise velocity component *v*, from NORTEK® "Vectrino" comparing measurements and right vectors direction), unlike classic unseeded LSPIV.

One could notice that all experiments were performed in total absence of suspending materials as tracer, so it is immediate to state that in presence of it (frequent condition in natural environment), the clustering-correction unseeded LSPIV could work even better combining recognition of materials convected by pure-current and surface deformations reflections due to capillary-gravity waves.

Finally, we would like to point out that LSPIV superficial velocity measurement is the most critical and innovative operation dealing with non-contact discharge monitoring. So, once a bathymetrical survey of a river (or channel) cross-section and a relation to pass from superficial to depth-averaged velocity are provided, the following step is the discharge calculation for permanent or even mobile installations.

Besides this, a simple novel procedure to orthorectify images has been presented allowing operator to remain in safety on the camera site.

**Acknowledgements**

The present study was funded by "Autorità di Bacino dei fiumi Isonzo, Tagliamento, Livenza, Piave, Brenta-Bacchiglione". The authors wish to express their gratitude to ARPAV ("Agenzia Regionale per la Prevenzione e Protezione Ambientale del Veneto – Servizio Idrografico Regionale", the hydrographic section of regional agency for environmental protection in Veneto, Italy) for providing comparing measurements on Bacchiglione River (Italy) and to prof. A. Adami (Protecno Srl, Italy) for precious comments.

**Table 1.** Technical specifications and custom parameters of NORTEK® "Vectrino", used during Experiments 1, 2 and 4 to compare water surface velocity measured by unseeded LSPIV.

| Parameter | Value |
|---|---|
| Sampling rate [Hz] | 1.0 |
| Nominal velocity range [m/s] | ± 1.0 |
| Transmit length [mm] | 2.4 |
| Sampling volume [mm] | 9.1 |

**Table 2.** Experiments 1 and 2, performed in the straight concrete-made rectangular-section channel. Velocity values from comparing instrument NORTEK® "Vectrino" measurements. σ is the standard deviation of measured velocity components. v is the streamwise velocity component, positive downstream (in Figures, from top to bottom), u is spanwise velocity component (in Figures, positive from left to right). Froude numbers for streamwise component of current are calculated assuming 1 m water depth.

| Experiment n° | Average measured $u$ velocity component in 60 s ± σ [m/s] | Average measured $v$ velocity component in 60 s ± σ [m/s] | Froude number of $v$ velocity component $Fr_v$ |
|---|---|---|---|
| *1* | 0.03 ± 0.02 | 0.42 ± 0.04 | 0.13 |
| *2* | 0.04 ± 0.01 | 0.89 ± 0.04 | 0.28 |

**Table 3.** Experiments 1 and 2, performed in the straight concrete-made rectangular-section channel. Space-averaged results of unseeded LSPIV measurements in 60 seconds. Velocity components are shown for (i) analysis without clustering and correction, i.e. classic LSPIV, ($\overline{U}_m$), (ii) pure current contribution after clustering (subscript "pc"), (iii) correction contribution according to c-g wave phase speed (subscript "corr") and (iv) average value from pure current and correction contributions.

| Experiment n° | $\overline{U}_{m,x}$ [m/s] | $\overline{U}_{m,y}$ [m/s] | $\overline{u}_{pc}$ [m/s] | $\overline{v}_{pc}$ [m/s] | $\overline{u}_{corr}$ [m/s] | $\overline{v}_{corr}$ [m/s] | $\overline{u}$ [m/s] | $\overline{v}$ [m/s] |
|---|---|---|---|---|---|---|---|---|
| *1* | -0.01 | 0.18 | -0.02 | 0.38 | 0.00 | 0.37 | -0.01 | 0.37 |
| *2* | 0.00 | 0.55 | 0.02 | 0.87 | -0.01 | 0.83 | 0.01 | 0.85 |

Table 4. Experiments 1 and 2, performed in the straight concrete-made rectangular-section channel. Results in the center of the channel, for comparison with NORTEK® "Vectrino" measurements (Table 2). LSPIV indicates results obtained from classic unseeded LSPIV, while ccLSPIV means clustering-correction unseeded LSPIV.

| Experiment n° | LSPIV $\bar{u}$ [m/s] | LSPIV $\bar{v}$ [m/s] | ccLSPIV $\bar{u}$ [m/s] | ccLSPIV $\bar{v}$ [m/s] |
|---|---|---|---|---|
| *1* | 0.00 | 0.22 | 0.00 | 0.44 |
| *2* | 0.04 | 0.61 | 0.02 | 0.91 |

Table 5. Experiment 4, performed in the straight concrete-made rectangular-section channel to test unseeded LSPIV performance when the mean current speed is comparable or even lower than the minimum c-g waves phase speed $\left|\vec{c}_{min}\right|$ = 0.23 m/s. Velocity values from comparing instrument NORTEK® "Vectrino" measurement. σ is the standard deviation of measured velocity components. v is the streamwise velocity component, positive downstream (in Figures, from top to bottom), u is spanwise velocity component (in Figures, positive from left to right). Froude numbers for streamwise component of current are calculated assuming 1 m water depth.

| Average measured *u* velocity component in 60 s ± σ [m/s] | Average measured *v* velocity component in 60 s ± σ [m/s] | Froude number of *v* velocity component $Fr_v$ |
|---|---|---|
| 0.00 ± 0.02 | 0.22 ± 0.02 | 0.07 |

Table 6. Experiment 4, performed in the straight concrete-made rectangular-section channel to test unseeded LSPIV performance when the mean current speed is comparable or even lower than the minimum c-g waves phase speed $\left|\vec{c}_{min}\right|$ = 0.23 m/s. Results in the center of the channel, for comparison with NORTEK® "Vectrino" measurements (Table 5). LSPIV indicates results obtained from classic unseeded LSPIV, while cLSPIV means correction unseeded LSPIV.

| LSPIV $\bar{u}$ [m/s] | LSPIV $\bar{v}$ [m/s] | cLSPIV $\bar{u}$ [m/s] | cLSPIV $\bar{v}$ [m/s] |
|---|---|---|---|
| 0.02 | 0.04 | 0.00 | 0.19 |

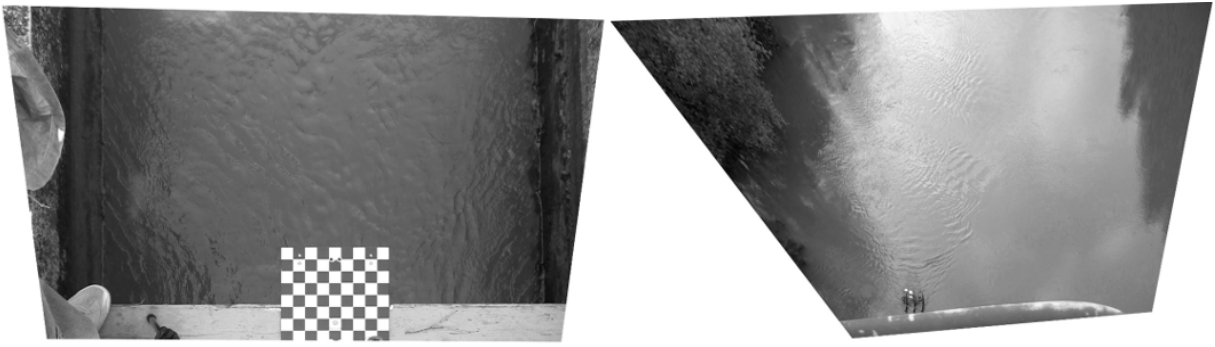

**Fig. 1. Examples of images used for unseeded LSPIV analysis. Left panel: concrete-made rectangular-section channel. Right panel: Bacchiglione River, Italy.**

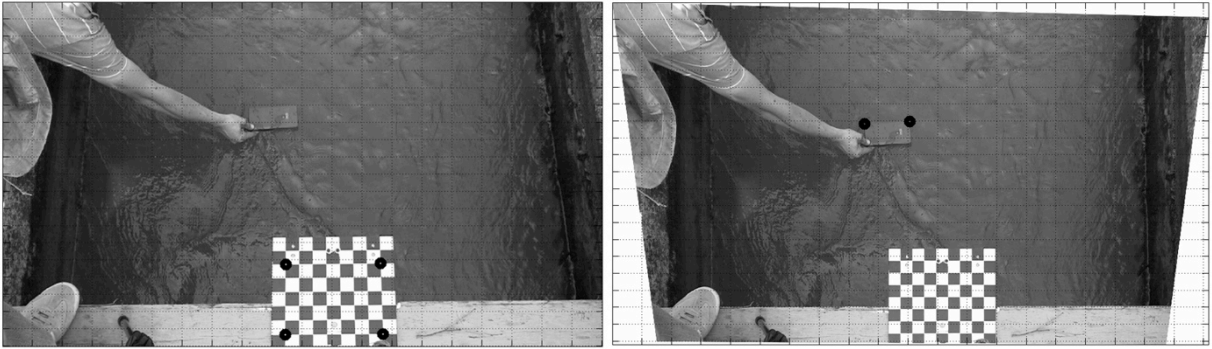

**Fig. 2. Orthorectification of images. Left panel: the chessboard is used to make a projective transformation to put, through at least four points coordinates (black filled circles), the images onto its reference frame, i.e. image x and y axes orthogonal to vertical direction. Right panel: an object of known dimensions (black filled circles) on the water surface is used to scale the images in order to put them onto water surface reference frame, parallel to chessboard reference frame**.

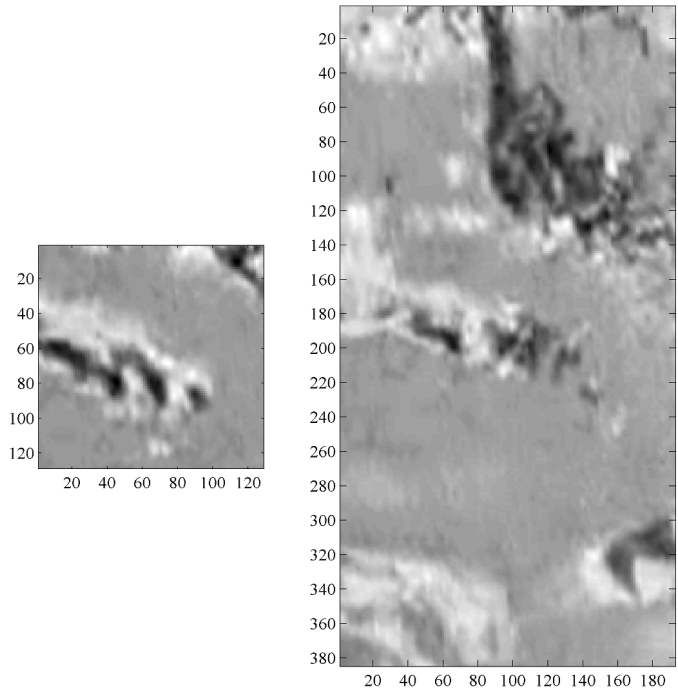

Fig. 3. Interrogation (IA) and Search (SA) Areas. Left panel: IA, 129 x 129 pixels. Right panel: SA, 193 x 385 pixels.

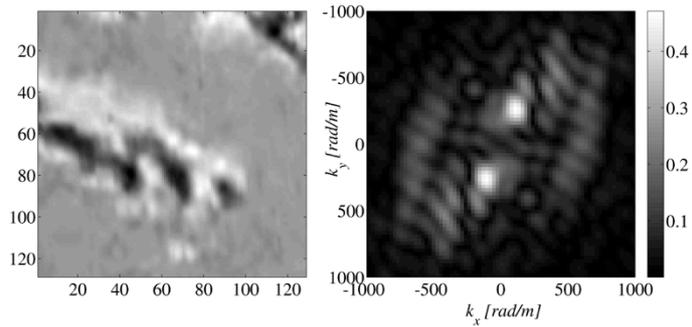

Fig. 4. 2D Fourier Transform of IA. Left panel: Interrogation Area (IA), where the crests of a c-g wave are visible. Image axes are in pixel. Right panel: absolute value of 2D Fourier Transform of IA; for clarity, only the central part is shown. The 2D spectrum is symmetric with respect to the center and in Figure example the peak wavenumber components are kx = 105.7 rad/m and ky = 266.4 rad/m, so direction of propagation is α ± Dα = 68.36° ± 5.28° with respect to x-axis. Nyquist wavenumber: [kx, ky]max = [6815.30 rad/m, 6815.30 rad/m]; spectral resolution: [kx, ky] = [26.41 rad/m, 26.64 rad/m].

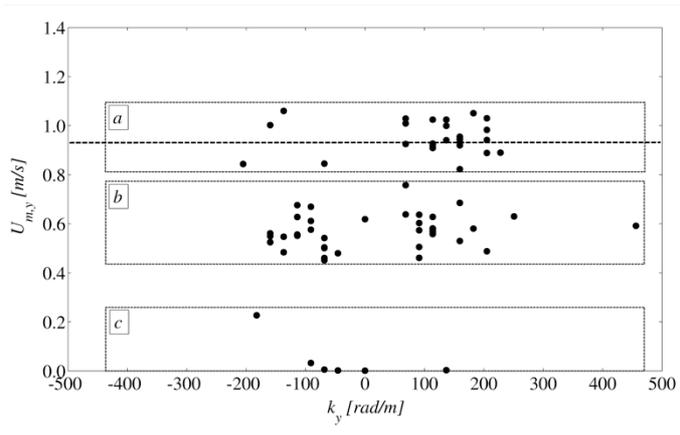

**Fig. 5.** Results of unseeded LSPIV. Example of streamwise velocity component $U_{m,y}$ measured at different time and at the same analysis point on the water surface, herein plotted versus streamwise wavenumber component $k_y$. Three candidate groups (Group a, Group b, Group c) of velocities are bounded with rectangles. Mean values of such groups are approximately 0.9 m/s, 0.6 m/s and 0.1 m/s. y axis is set parallel to the expected mean flow direction. In Figure, the mean flow velocity value, measured in the same test with a reference instrument, is also shown with a dashed line.

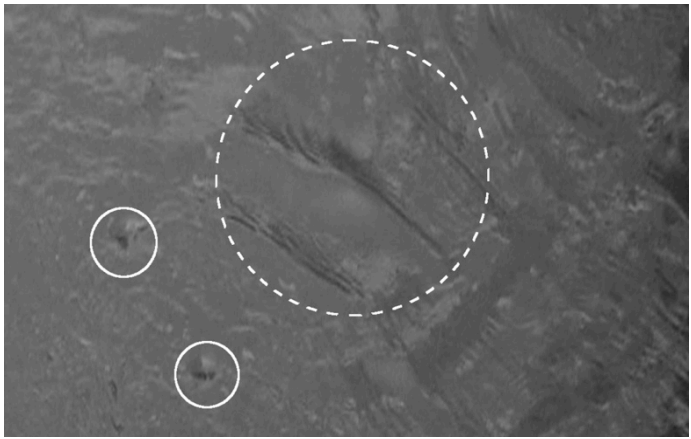

**Fig. 6.** Examples of moving structures on water surface. Circled with continuous line: vortexes convected downstream by current; circled with dashed line: a sequence of c-g wave crests.

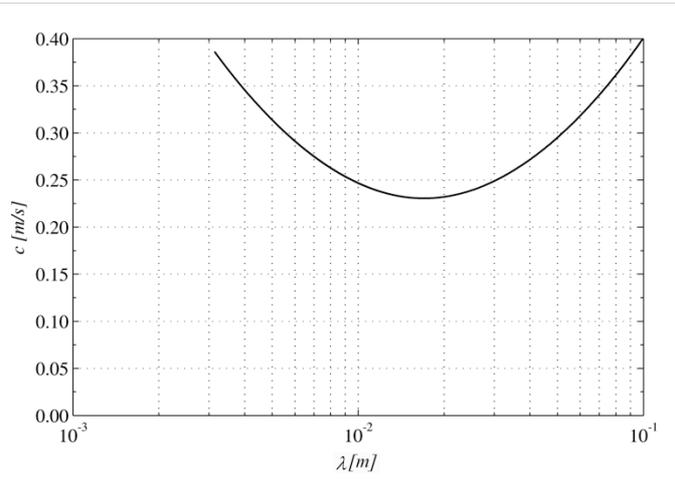

Fig. 7. C-g waves phase speed (10), magnitude. The function shows a minimum at λ = 0.017 m, corresponding to $|\vec{c}|$ = 0.23 m/s.

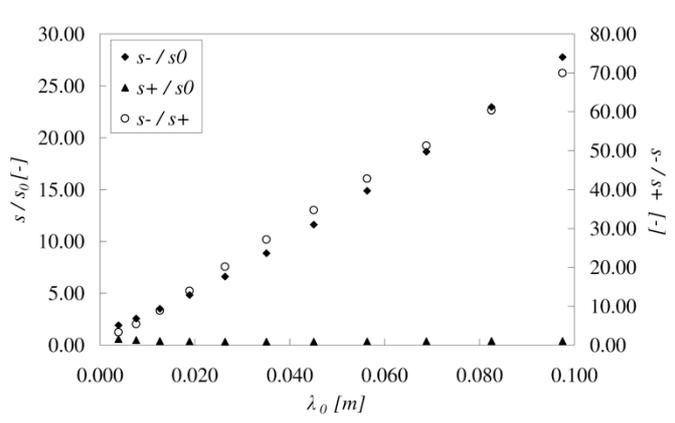

Fig. 8. Example of c-g wave slope changing for different wavelength in still water l0 and same still water wave height H0 = 0.005 m. In Figure three slope ratios are plotted: s+ / s0, s- / s0, s- / s+, where s0 is slope in still water, s+ is wave slope (H+/λ+) with current in the same wave propagation sense and direction (assuming $|\vec{U}|$ = + 0.20 m/s, in this example) and s- is wave slope (H-/λ-) with current flowing opposed to wave propagation (assuming $|\vec{U}|$ = - 0.20 m/s, in this example). s- / s+ ratio shows that c-g waves slope under the effect of current steepening is much higher than under the effect of current flattening.

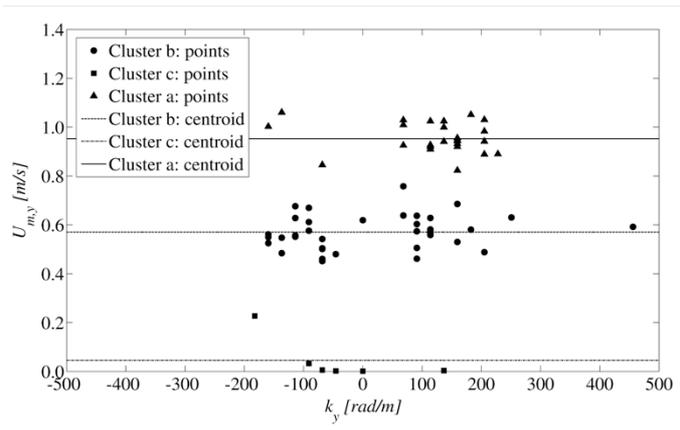

Fig. 9. Example of clustering of streamwise velocity component $U_{m,y}$ measured at different time and at the same analysis point. Experimental points are grouped together in 3 clusters and their centroids are shown.

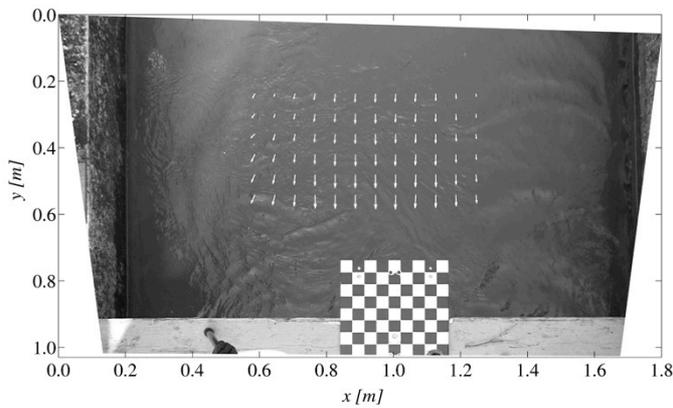

Fig. 10. Experiment 1: 60 s-mean 2D velocity vector field obtained applying classic unseeded LSPIV (without clustering and correction).

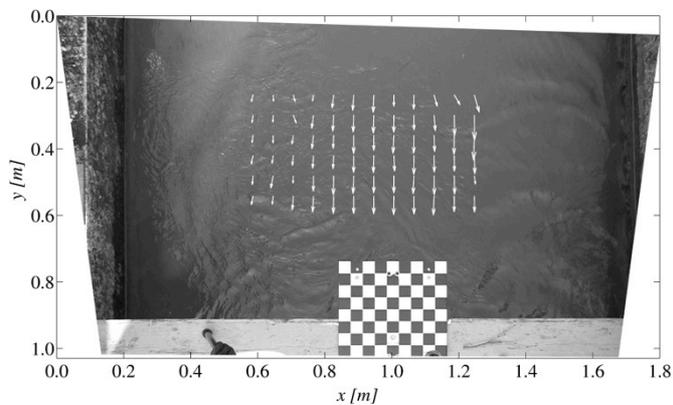

Fig. 11. Experiment 1: 60 s-mean 2D velocity vector field obtained applying clustering-correction unseeded LSPIV, in accordance with (11).

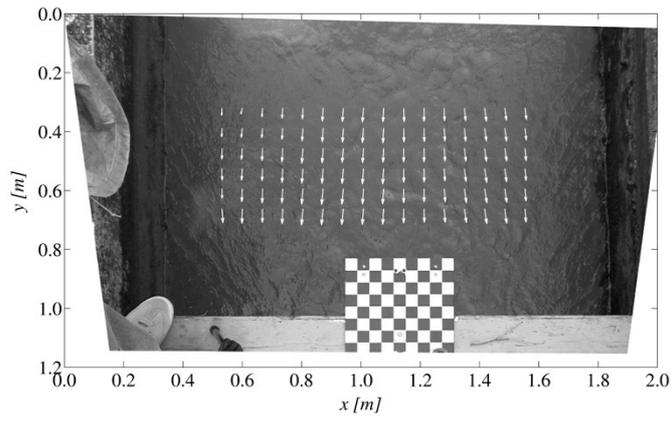

**Fig. 12. Experiment 2: 60 s-mean 2D velocity vector field obtained applying classic unseeded LSPIV (without clustering and correction).**

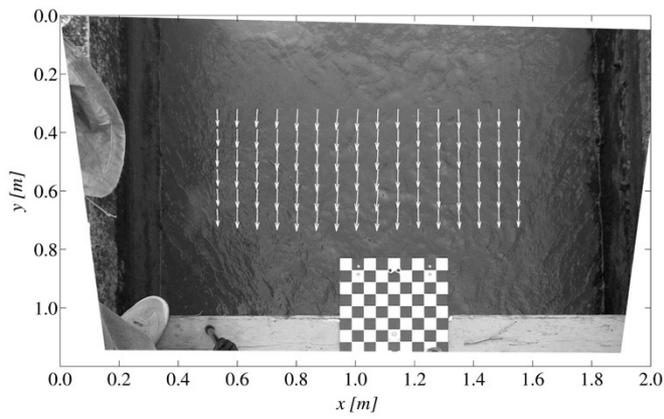

**Fig. 13. Experiment 2: 60 s-mean 2D velocity vector field obtained applying clustering-correction unseeded LSPIV, in accordance with (11).**

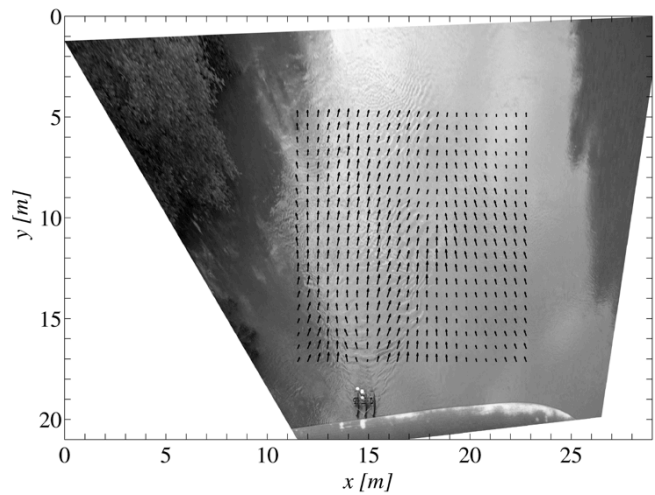

**Fig. 14. Experiment 3: 40 s-mean 2D velocity vector field obtained applying classic unseeded LSPIV (without clustering and correction).**

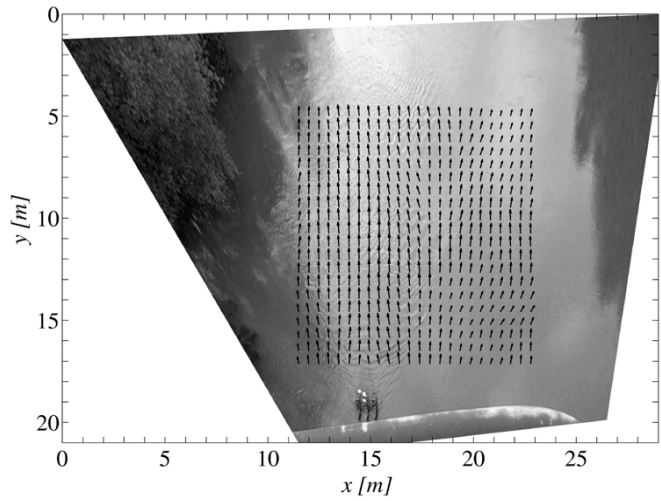

**Fig. 15. Experiment 3: 40 s-mean 2D velocity vector field obtained applying clustering-correction unseeded LSPIV, in accordance with (11).**

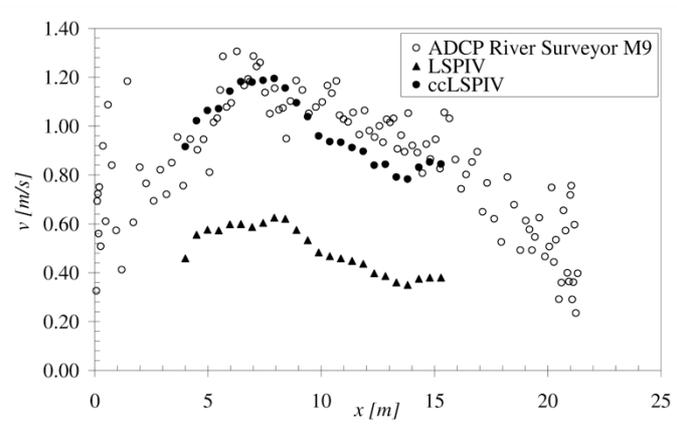

Fig. 16. Experiment 3: results from classic LSPIV (in Figure, LSPIV), clustering-correction LSPIV (in Figure, ccLSPIV) and reference instrument SONTEK® "River Surveyor M9" measurements, taken along a cross-section of Bacchiglione River. Classic and clustering-correction LSPIVs represent velocity component v averaged over all points of same x coordinate of Fig. 14 e Fig. 15. Froude number of ADCP measured velocity component v is Frv = 0.17 assuming a 2.50 m average water depth over the river cross-section. LSPIVs measurements are time-averaged over 40 s, while "River Surveyor M9" ones are instantaneous at different x coordinates.

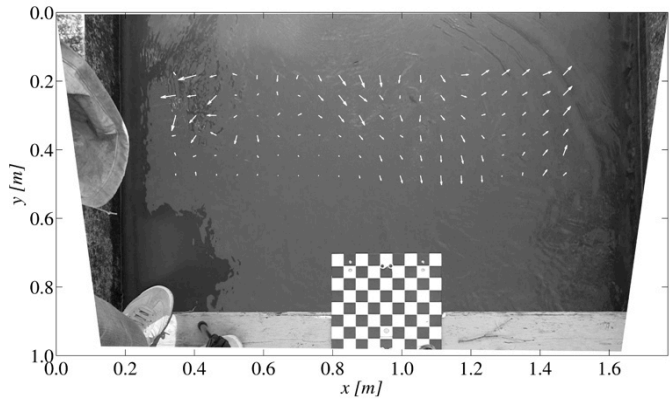

Fig. 17. Experiment 4, performed in the straight concrete-made rectangular-section channel to test unseeded LSPIV performance when the mean current speed is comparable or even lower than the minimum c-g waves phase speed $\left|\vec{c}_{min}\right|$ = 0.23 m/s. 60 s-mean 2D velocity vector field obtained applying classic unseeded LSPIV (without correction).

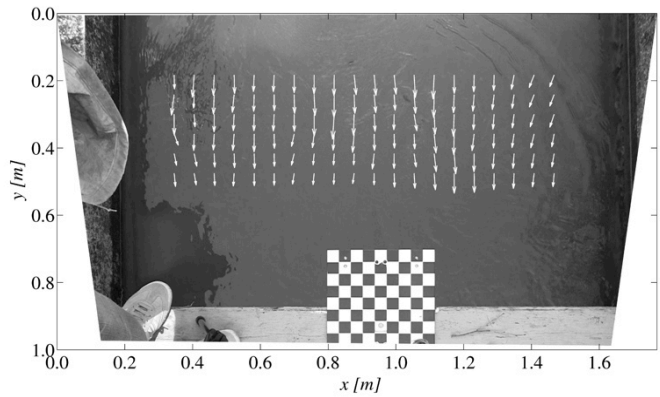

Fig. 18. Experiment 4, performed in the straight concrete-made rectangular-section channel to test unseeded LSPIV performance when the mean current speed is comparable or even lower than the minimum c-g waves phase speed $\left|\vec{c}_{min}\right|$ = 0.23 m/s. 60 s-mean 2D velocity vector field obtained applying correction unseeded LSPIV, in accordance with (11).

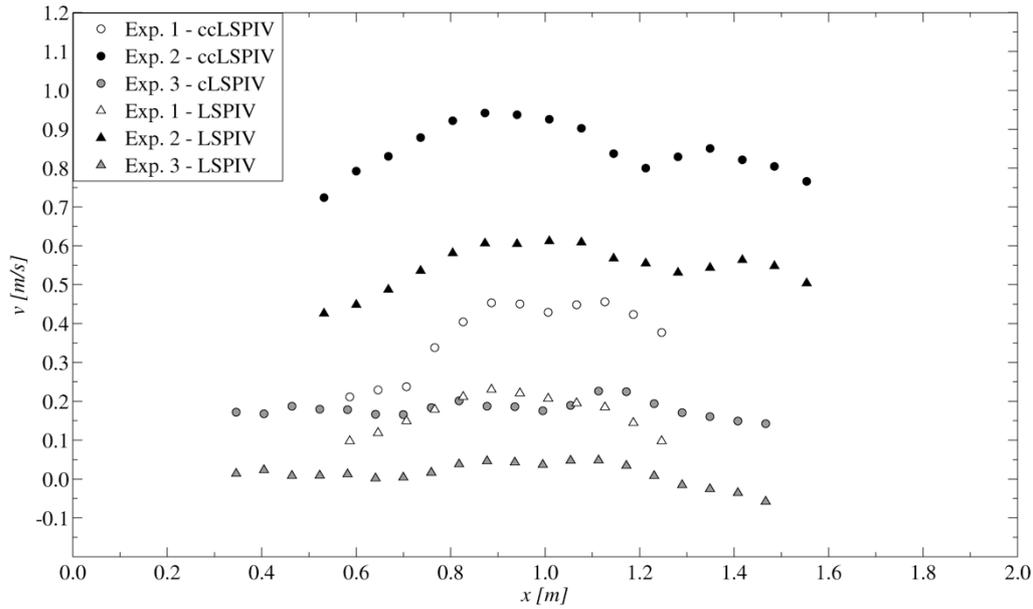

Fig. 19. Experiments (abbreviated Exp.) 1, 2 and 4 in the straight concrete-made rectangular-section channel. 60 s-mean streamwise averaged v profiles along a cross-section for classic (LSPIV) and clustering-correction (ccLSPIV, Experiments 1 and 2) or only correction (cLSPIV, Experiment 4) unseeded LSPIVs.

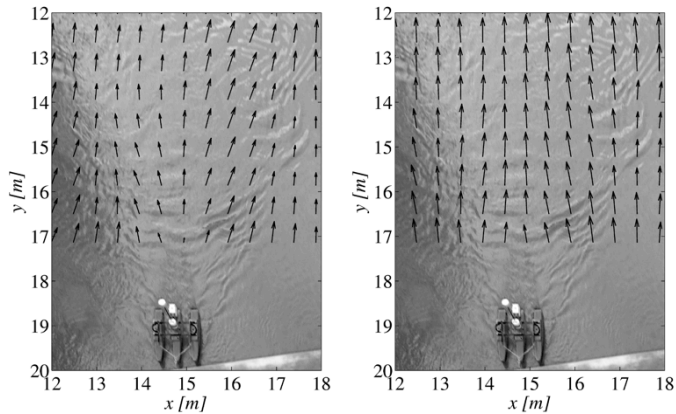

Fig. 20. Experiment 3. Effect of correction accounting for c-g waves phase speed on velocity vectors field. Left panel: unseeded LSPIV velocity vectors composed of pure-current contribution and the one of ship waves in the trimaran wake. Right panel: after velocity clustering and correction, vectors represent the only pure current contribution.

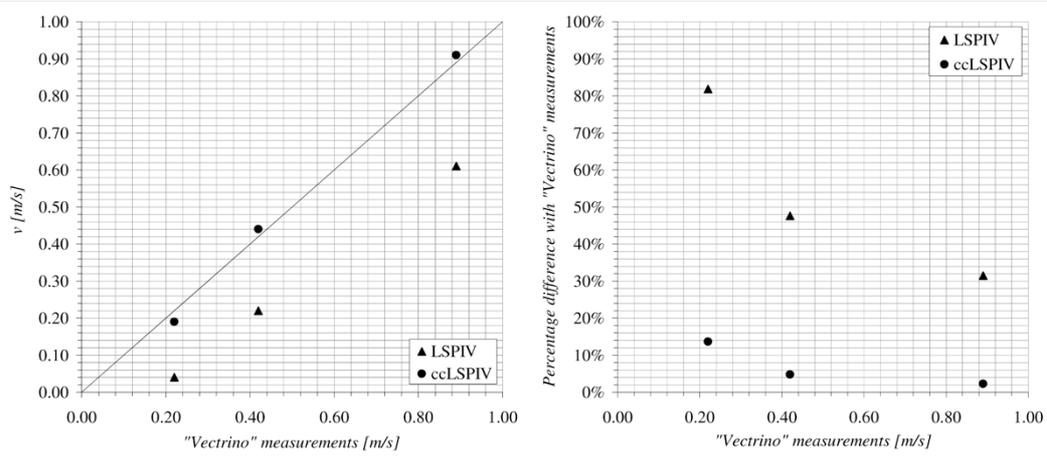

Fig. 21. Experiments in the straight concrete-made rectangular-section channel (Experiment 1, 2 and 4): measurements of velocity component v in front of "Vectrino", in the center of the channel. Importance of taking into account c-g waves phase speed in unseeded LSPIV measurements. LSPIV indicates results obtained from classic LSPIV, while ccLSPIV means clustering-correction LSPIV (for Experiment 4, correction LSPIV). Left panel: LSPIV and ccLSPIV results. Right panel: percentage difference of velocity component v from classic unseeded LSPIV and clustering-correction unseeded LSPIV with respect to velocity component v from "Vectrino" measurements. As velocity increases ("Vectrino" measurements), percentage difference between "Vectrino" and unseeded LSPIV measurements (both classic and clustering-correction) decreases. However, clustering-correction unseeded LSPIV results are clearly more accurate than classic unseeded LSPIV ones.

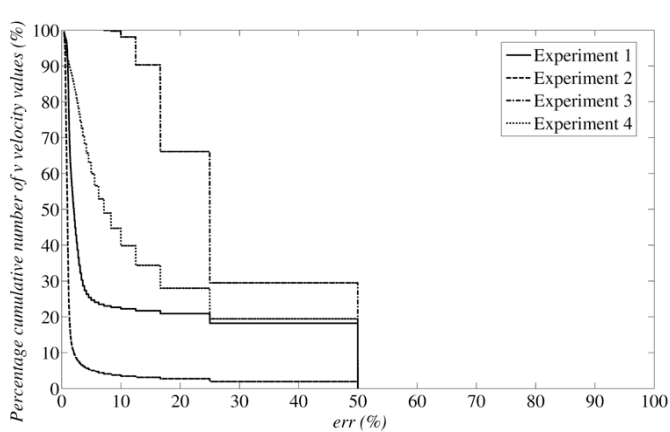

Fig. 22. PIV-related exceeded error err due to integer ds displacement calculation for the experiments reported as err = es / ds, with es uncertainty in pixel displacement calculation (= 0.5 pixels). Neglecting sub-pixel result (whose uncertainty is unknown) maximum exceeded error is 50%. In Figure, every potential error value is related to the cumulative percentage of streamwise velocity component v values measured by LSPIV. Experiment 3 curve is stepped since integer displacement is strongly quantized, because of distance from camera to water surface. Moreover, dependence of err from mean current speed magnitude is clearly visible comparing Experiments 1, 2 and 4 curves.